\documentclass{aa}  

\usepackage{graphicx}
\usepackage{natbib}
\usepackage{txfonts}
\usepackage{amsmath}
\usepackage{multirow}
\usepackage{float}
\usepackage[colorlinks=true,     linkcolor=blue, citecolor=blue, filecolor=blue, urlcolor=blue]{hyperref}
\begin{document} 
\nolinenumbers

   \title{Extreme FeLoBAL outflow in the VLT/UVES spectrum of quasar SDSS J1321-0041}

   \author{Doyee Byun
          %\inst{1}
          \and
          Nahum Arav%\inst{1}
          \and
          Mayank Sharma%\inst{1}
          \and
          Maryam Dehghanian%\inst{1}
          \and
          Gwen Walker%\inst{1}
          }

   \institute{Department of Physics, Virginia Tech,
              Blacksburg, VA, USA\\
              \email{dbyun@vt.edu}
             }

   %\date{Received September 15, 1996; accepted March 16, 1997}

% \abstract{}{}{}{}{} 
% 5 {} token are mandatory
 
  \abstract
   {Quasar outflows are often analyzed to determine their ability to contribute to active galactic nucleus (AGN) feedback. We identified a broad absorption line (BAL) outflow in the VLT/UVES spectrum of the quasar SDSS J1321-0041. The outflow shows troughs from \ion{Fe}{ii}, and is thus categorized as an FeLoBAL. This outfow is unusual among the population of FeLoBAL outflows, as it displays \ion{C}{ii} and \ion{Si}{ii} BALs.}
   {Outflow systems require a kinetic luminosity above $\sim0.5\%$ of the quasar's luminosity to contribute to AGN feedback. For this reason, we analyzed the spectrum of J1321-0041 to determine the outflow's kinetic luminosity, as well as the quasar's bolometric luminosity.}
   {We measured the ionic column densities from the absorption troughs in the spectrum and determined the Hydrogen column density and ionization parameter using those column densities as our constraints. We also determined the electron number density, $n_e$, based on the ratios between the excited-state and resonance-state column densities of \ion{Fe}{ii} and \ion{Si}{ii}. This allowed us to find the distance of the outflow from its central source, as well as its kinetic luminosity.}
   {We determined the kinetic luminosity of the outflow to be $8.4^{+13.7}_{-5.4}\times 10^{45}\text{ erg s}^{-1}$ and the quasar's bolometric luminosity to be $1.72\pm0.13\times10^{47}\text{ erg s}^{-1}$, resulting in a ratio of $\dot{E}_k/L_{Bol}=4.8^{+8.0}_{-3.1}\%$. We conclude that this outflow has a sufficiently high kinetic luminosity to contribute to AGN feedback.}
   {}

   \keywords{ Galaxies: active -- Quasars: absorption lines
                 -- Quasars: individual: SDSS J132139.86-004151.9
               }

   \maketitle
%
%-------------------------------------------------------------------

\section{Introduction}

Active galactic nucleus (AGN) feedback is a process in which an AGN affects the evolution of its host galaxy, including the regulation of the star formation rate and the correlation between the black hole and host galaxy mass \citep[e.g.,][]{1998A&A...331L...1S,2009ApJ...699...89C,2016MNRAS.455.1211K}.  In quasars, AGN feedback is often attributed to outflows, which are found in $\lesssim 40\%$ quasar spectra as blueshifted absorption troughs \citep[e.g.,][]{2003AJ....125.1784H,2008ApJ...672..108D,2008MNRAS.386.1426K,2021ApJ...919..122V,2022SciA....8.3291H}. Outflow systems require a kinetic luminosity ($\dot{E}_k$) above $\sim0.5\%$ of the quasar's luminosity \citep{2010MNRAS.401....7H}, which \citet{2020MNRAS.499.1522M} interpreted to be the Eddington luminosity ($L_{Edd}$). Outflow analyses have been conducted in past works, reporting quasar outflows with a sufficient value of $\dot{E}_k$  \citep[e.g.,][]{2015MNRAS.450.1085C,2018ApJ...866....7L,2020ApJS..247...39M,10.1093/mnras/stac2638,2022Byun,Q0254_Byun,2022MNRAS.516.3778W}.\par
The process of finding $\dot{E}_k$ involves measuring the ionization parameter ($U_H$) and electron number density ($n_e$) of the outflow, which leads to the distance from the central source ($R$) and mass outflow rate ($\dot{M}$) \citep{2012ApJ...758...69B}. For the analysis of ionized outflow, the spectral synthesis code \textsc{Cloudy} \citep{2017RMxAA..53..385F} can be used to compare measured ionic column densities with simulated values from models created from a range of $U_H$ and hydrogen column density ($N_H$) values \citep[e.g.,][]{2020ApJS..247...39M,10.1093/mnras/stac2194,Q0254_Byun,2022MNRAS.516.3778W}.\par
A category of BAL quasars is known as iron low-ionized BAL (FeLoBAL) quasars, due to the identification of \ion{Fe}{ii} absorption troughs in their spectra. Recent studies of FeLoBALs include (but are not limited to) the study of a powerful FeLoBAL outflow in the quasar SDSS J135246.37+423923.5 \citep{2020ApJ...891...53C}; the analysis of the FeLoBAL quasar Q0059-2735, whose outflow has shown signs of broad \ion{Si}{ii} absorption \citep{2021MNRAS.506.2725X}; and a systematic study of the properties of FeLoBAL quasars using spectral synthesis code SimBAL \citep[][]{2022ApJ...936..110C,2022ApJ...937...74C,2022ApJ...935...92L}.\par
We present the analysis of the UVES spectrum of the quasar SDSS J132139.86-004151.9 (hereafter, J1321-0041), which was retrieved from the Spectral Quasar Absorption Database (SQUAD) published by \citet{Murphy2019}. The data from SQUAD have 20 times higher spectral resolution than SDSS spectra,  therefore lending themselves to a more detailed analysis. We conducted our analysis through the method described above in order to find $\dot{E}_k$ and to determine the outflow's potential ability to contribute to AGN feedback. Similar analyses of quasar outflows have been conducted in past works using spectra from SQUAD \citep[e.g.,][]{2022Byun,10.1093/mnras/stac2638,2022MNRAS.516.3778W}.\par
This paper is structured as follows. Section~\ref{sec:observation} describes the observation of J1321-0041, as well as our data acquisition process. Section~\ref{sec:analysis} describes the process through which we have found the outflow's ionic column densities, $N_H$, $U_H$, and $n_e$ values. Section~\ref{sec:results} presents the resulting energetics parameters $R$ and $\dot{E}_k$. Section~\ref{sec:discussion} discusses the outflow's potential ability to contribute to AGN feedback and compares it to other outflows that have been analyzed in the past. Section~\ref{sec:conclusion} summarizes and concludes the paper. For our analysis, we adopted a cosmology of $h=0.696, \Omega_m=0.286$, and $\Omega_\Lambda=0.714$ \citep{Bennett_2014}.\par
    \begin{figure*}
            \includegraphics[width=17cm]{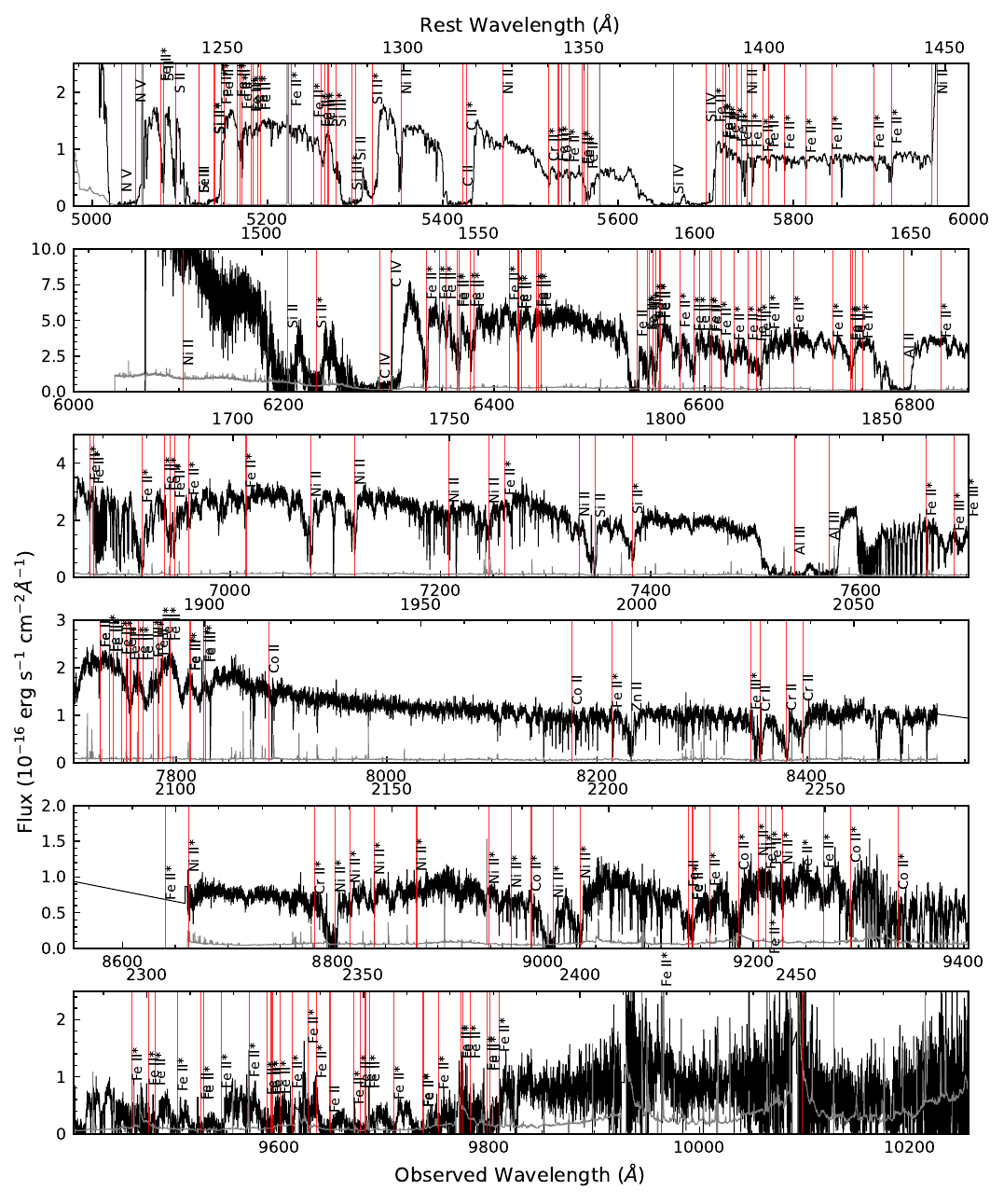}
      \caption{UVES spectrum of J1321-0041. Located absorption troughs and expected locations of absorption troughs from an outflow system, at $v\approx-4100$ km s$^{-1}$, are marked with red vertical lines. The flux has been scaled to match the flux of the SDSS spectrum at observed wavelength $\lambda=6850 \AA$.
              }
         \label{fig:spectrum}
    \end{figure*}
\section{Observations and data acquisition}
\label{sec:observation}
\begin{figure}
   \centering
   \resizebox{\hsize}{!}
        {\includegraphics{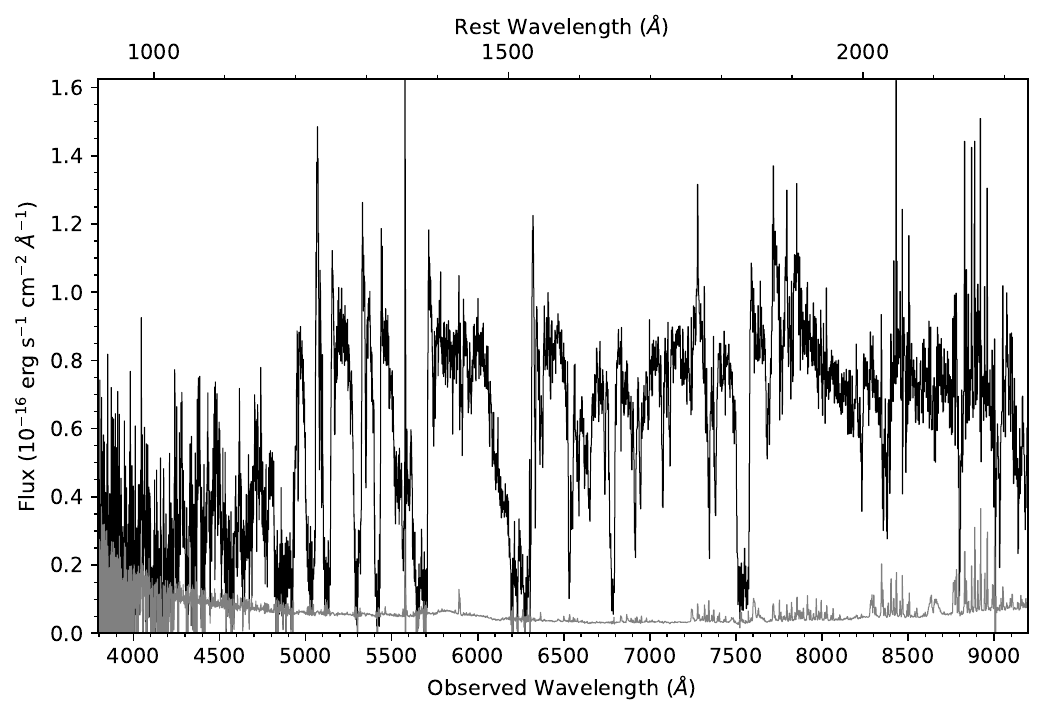}}
      \caption{SDSS spectrum of J1321-0041 that was used to scale the continuum flux of the UVES spectrum and to calculate the quasar's bolometric luminosity. The flux is plotted in black and the error is plotted in gray.
              }
         \label{fig:sdss}
   \end{figure}

J1321-0041 \citep[J2000: RA=13:21:39.86, DEC: -00:41:51.9, z=3.119,][]{2010MNRAS.405.2302H} was observed with VLT/UVES on February 13, 2008, with a total exposure time of 22,800 seconds, as part of program 080.B-0445(A). \citet{Murphy2019} normalized the spectrum by the quasar's continuum and emission and added it to SQUAD. The quasar was also observed on 4 February 2000, 1 May 2000, and 16 March 2001 as part of the Sloan Digital Sky Survey (SDSS), with little to no time variability between observations. We co-added the SDSS spectra to scale the flux of the UVES spectrum shown in Figure~\ref{fig:spectrum} and to find the bolometric luminosity of the quasar. The co-added SDSS spectrum can be seen in Figure~\ref{fig:sdss}.\par
We have identified broad absorption line (BAL) outflow traveling at $v\approx -4100\text{ km s}^{-1}$ with troughs of multiple ionic transitions in the UVES spectrum of J1321-0041, including \ion{C}{iv}, \ion{Si}{iv}, \ion{Si}{ii}, and \ion{Fe}{ii}. The velocity of the outflow was determined based on the deepest part of the \ion{Si}{ii} $\lambda1808$ absorption. The detection of \ion{Fe}{ii} absorption features puts this outflow in the class of FeLoBALs. The outflow is unusual, as its \ion{C}{ii} trough by itself adheres to the definition of a BAL \citep{1991ApJ...373...23W}, as its width is $\sim2300\text{ km s}^{-1}$. Only relatively few such objects are described in the literature\citep{2021MNRAS.506.2725X,2022ApJ...937...74C}. The presence of multiple energy states of \ion{Fe}{ii} and \ion{Si}{ii} have enabled us to estimate the outflow's $n_e$, $R$, and $\dot{M}$ values, and by extension, $\dot{E}_k$. The \ion{Fe}{ii}* transition lines used for this paper's analysis, and their associated energies, are shown in Table~\ref{table:iron}.
\begingroup
\setlength{\tabcolsep}{10pt}
\renewcommand{\arraystretch}{1.5}
\begin{table}
\caption{\ion{Fe}{ii}* lines that were used for the $n_e$ analysis.}
\centering
\begin{tabular}{lcc}
\hline\hline
$\lambda$ (\AA)&Energy (cm$^{-1}$)&$\log{n_{crit}}$ [cm$^{-3}$]\\
\hline
1628.160 &667 & 4.4\\
1623.093 &2430 & 4.5\\
1570.245 &2837 & 4.6\\
1726.393 &3117 & 4.6\\
1562.270 &7955 & 4.7\\
\hline
\end{tabular}
\tablefoot{Columns show transition wavelengths, energies, and critical densities, $\log{n_{crit}}$.}
\label{table:iron}
\end{table}
\endgroup
\section{Analysis}
\label{sec:analysis}
\subsection{Ionic column densities}
\label{subsec:coldensity}
\begin{figure*}
    \resizebox{\hsize}{!}
            {\includegraphics{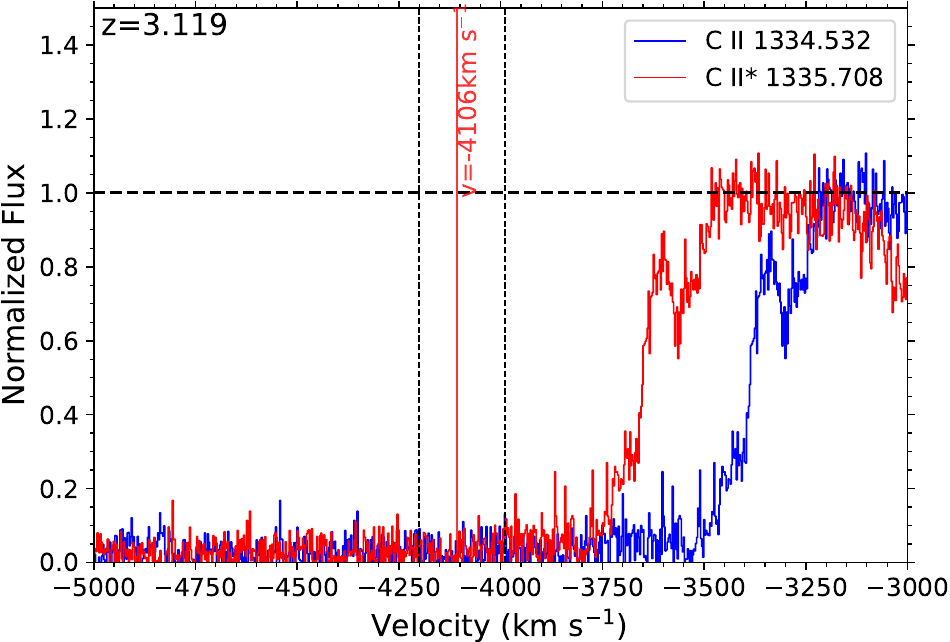}
            \includegraphics{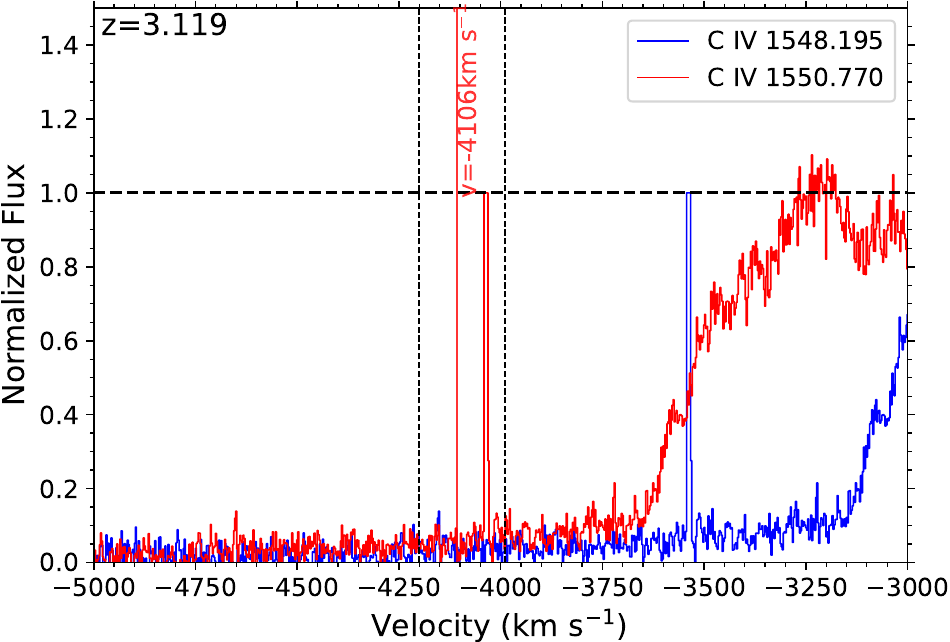}
            \includegraphics{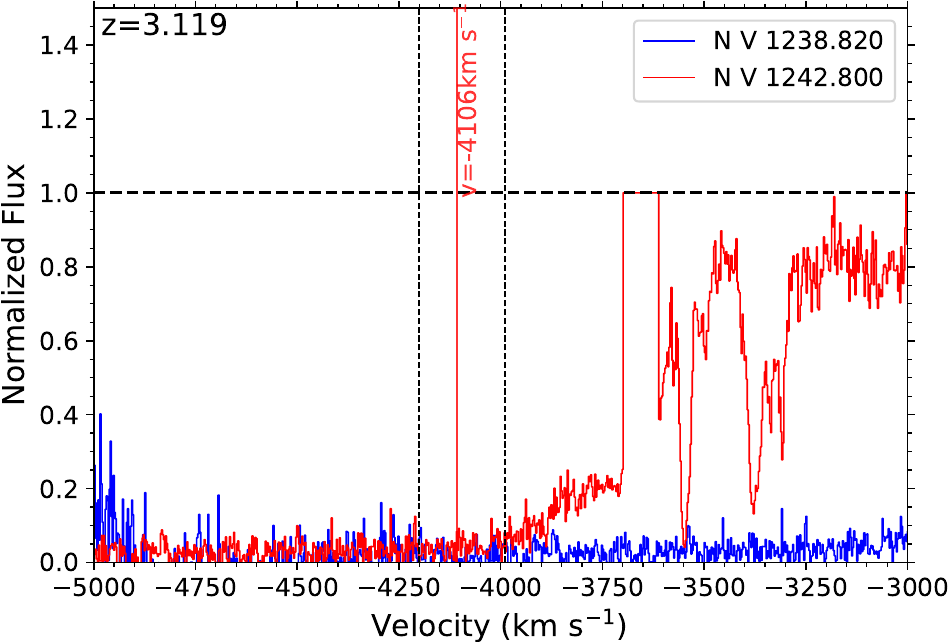}}
    \resizebox{\hsize}{!}
            {\includegraphics{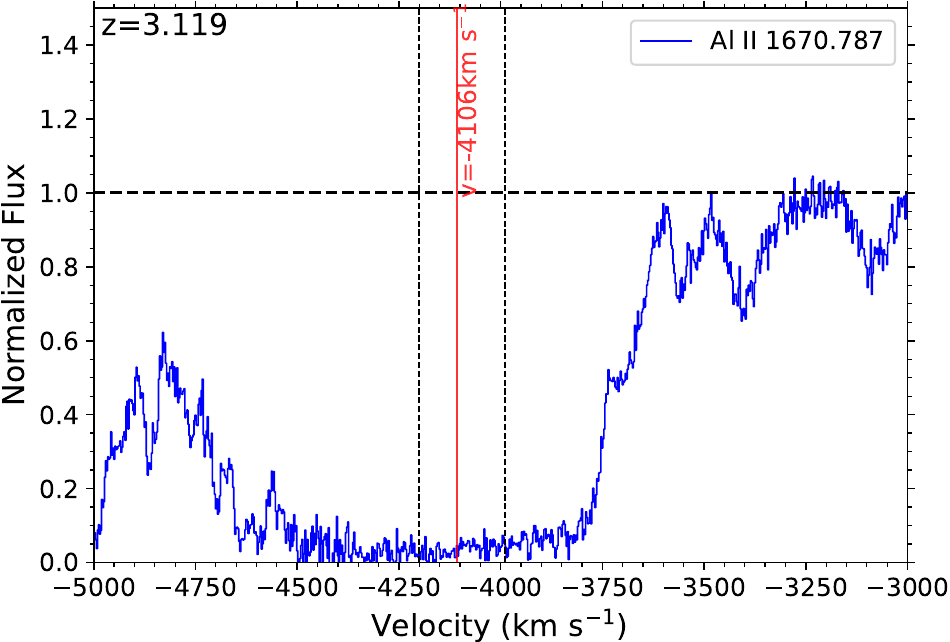}
            \includegraphics{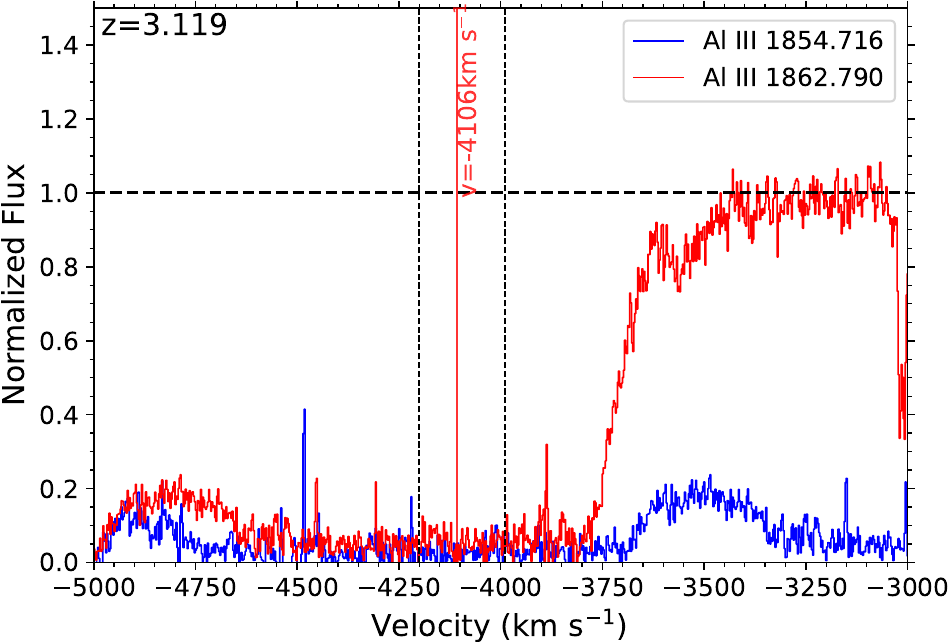}
            \includegraphics{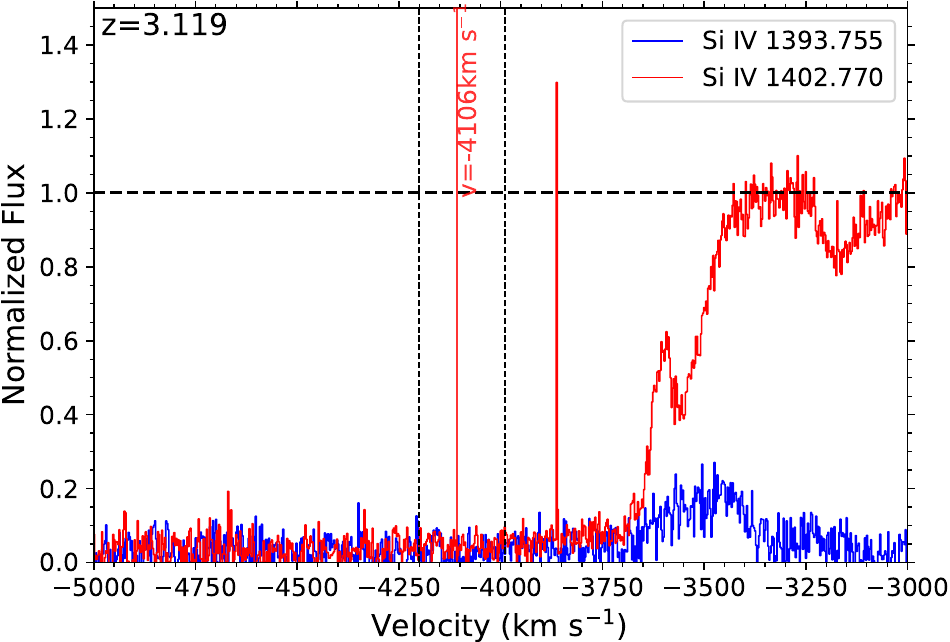}}
            \centering
        \includegraphics[width=7cm]{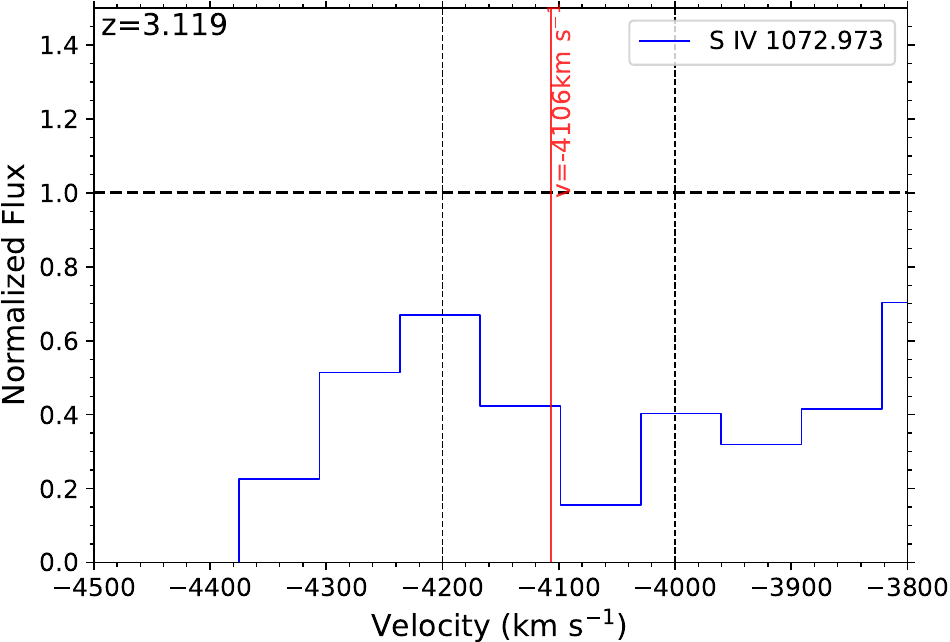}
        \includegraphics[width=7cm]{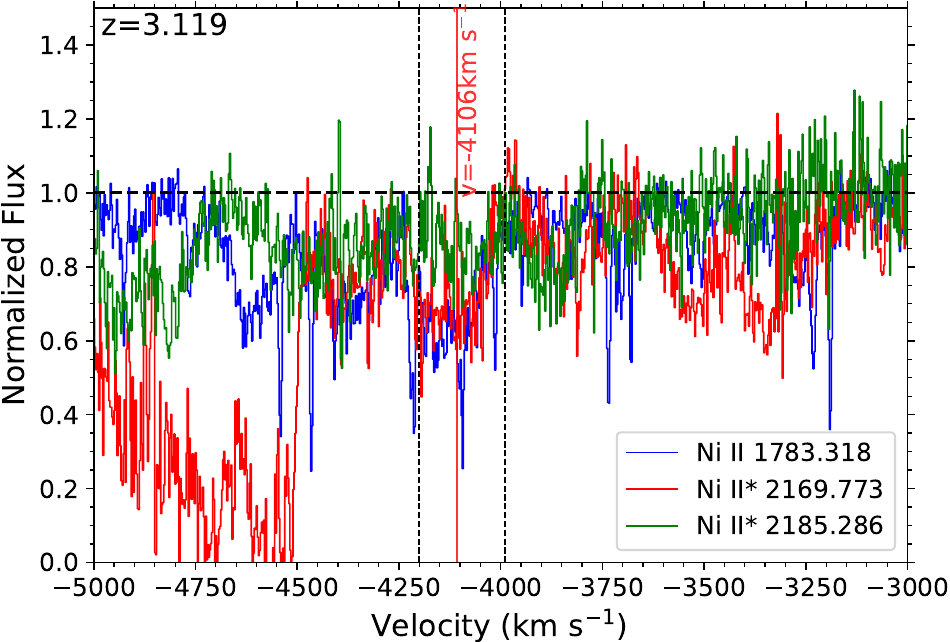}\\
        \includegraphics[width=7cm]{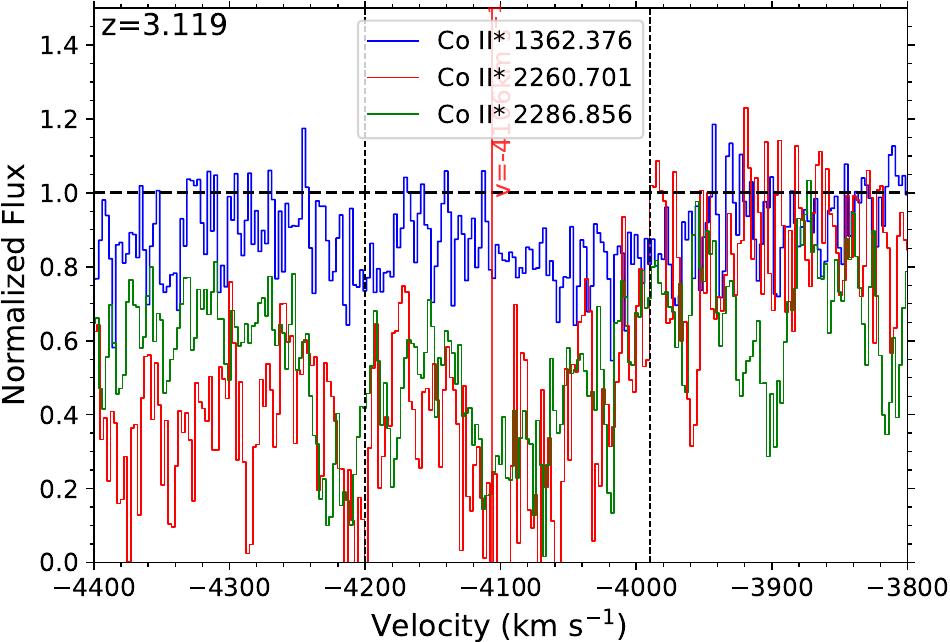}
        \includegraphics[width=7cm]{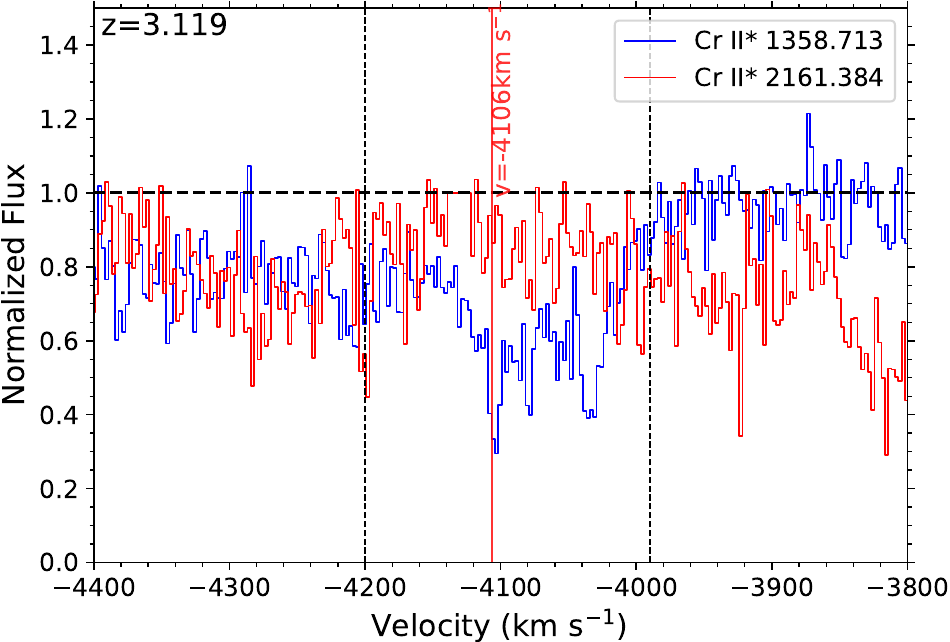}\\
    \resizebox{\hsize}{!}
            {\includegraphics{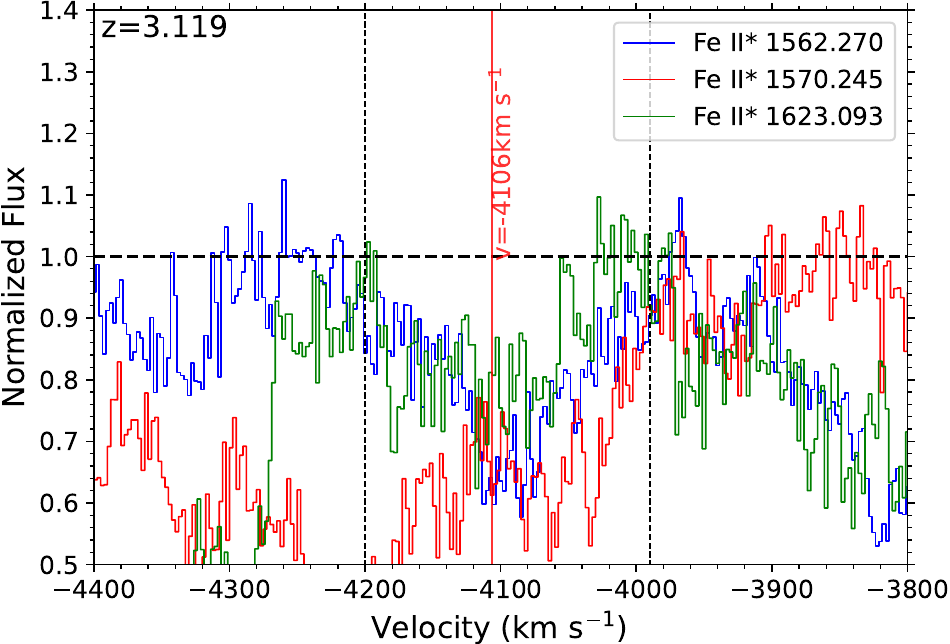}
            \includegraphics{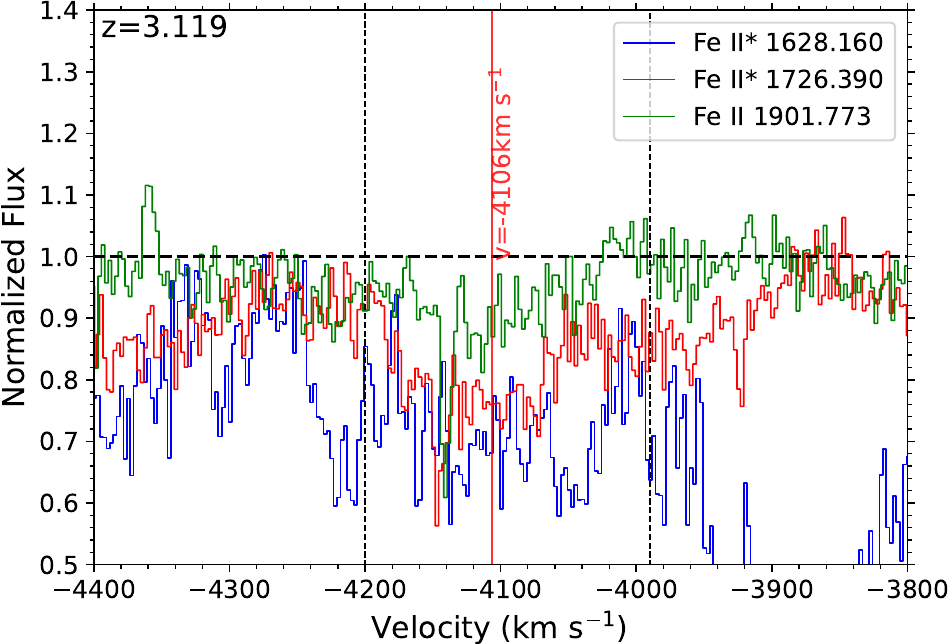}
            \includegraphics{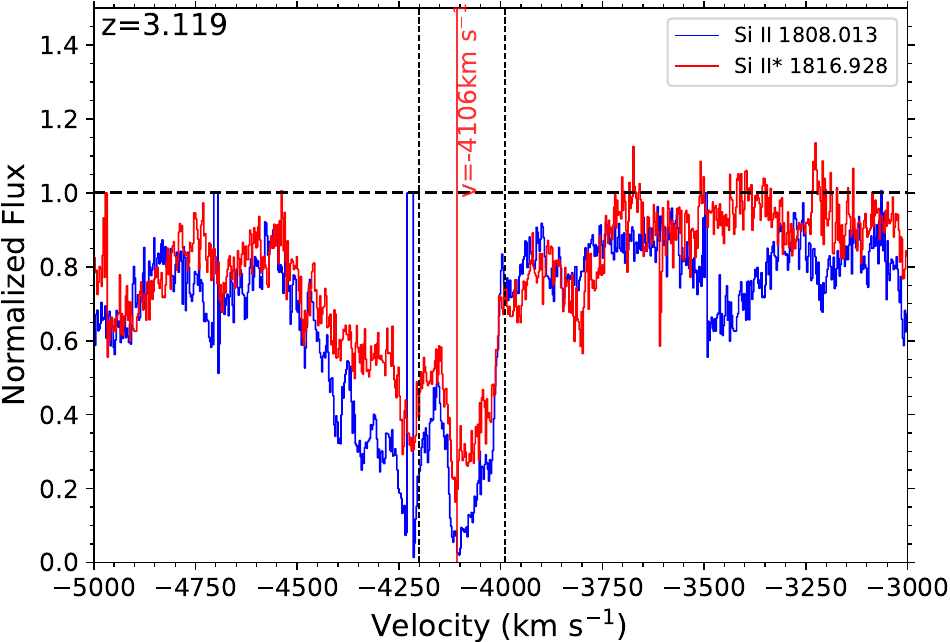}}
      \caption{Normalized spectrum of J1321-0041, converted into velocity space. Troughs of individual ionic transitions are color coded, and the integration ranges used for column density calculations are marked by vertical dotted lines. The continuum is represented by a horizontal dashed line. All spectra shown are UVES spectra, except for that of the \ion{S}{iv} absorption, which is based on the SDSS spectra due to wavelength range limitations. We note the stark contrast in the signal-to-noise ratio. The bottom three panels show the \ion{Fe}{ii} and \ion{Si}{ii} absorption, which were used to estimate $n_e$.
              }
         \label{fig:vcut}
    \end{figure*}

The first step of our analysis was to find the column densities ($N_{ion}$) of the ions found in the outflow system. We converted the normalized UVES spectrum from wavelength space to velocity space using the systemic redshift of the quasar (see Figure~\ref{fig:vcut}), and measured the ionic column densities assuming an apparent optical depth (AOD) of a uniform and homogeneously covering outflow \citep{1978ppim.book.....S,Savage1991}.\par
The AOD method relates the intensity and optical depth as follows:
\begin{equation}
    I(\lambda)=I_0 (\lambda)e^{-\tau(\lambda)}
,\end{equation}
where $I(\lambda)$ is intensity as a function of wavelength, $I_0(\lambda)$ is what the intensity would be without absorption, and $\tau(\lambda)$ is the optical depth. Normalizing a spectrum makes the process of finding the optical depth relatively easier, as it becomes a matter of relating the normalized intensity $I(\lambda)/I_0(\lambda)$ to $e^{-\tau(\lambda)}$. Finding the optical depth allows us to find the column density, as they are related as follows \citep{1978ppim.book.....S}:
\begin{equation}
    \tau(v)=\frac{\pi e^2}{m_e c}f\lambda N(v)
,\end{equation}
where $e$ is elementary charge (C), $m_e$ is electron mass (kg), $f$ is the oscillator strength of the ionic transition (unitless), $\lambda$ is the wavelength of the transition (\AA), and $N(v)$ is column density per unit velocity (cm$^{-2}$/km s$^{-1}$). The column density of an ion can be found by integrating $N(v)$ over the velocity range of the absorption trough. While this is a simple and straightforward method to estimating the ionic column density, it is limited to finding the lower limits when measuring column densities from saturated lines.\par
We identified the velocity range of the \ion{Fe}{ii} absorption to be $-4200\text{ km s}^{-1} \lesssim v \lesssim -4000\text{ km s}^{-1}$ (see bottom panel of Figure~\ref{fig:vcut}) and integrated over that range to find the ionic column densities of the outflow, as shown in Table~\ref{table:columndensity}. The red boundary was determined based on the red wing of the deepest absorption trough of \ion{Si}{ii}. The integration range was kept consistent among the different ions to allow direct comparison of different column densities. The \ion{Cr}{ii} lines that we identified are those of \ion{Cr}{ii}* $\lambda1358.71$ and $\lambda2161.38$. Despite the oscillator strength of the latter being three orders of magnitude larger ($f=0.014$ vs. $2.3\times10^{-5}$), the depth of the former trough is deeper, suggesting that the absorption features may be contaminated (See Figure~\ref{fig:vcut}). Thus, we determined that we would be unable to reliably measure the column density of \ion{Cr}{ii}. We have identified two visible \ion{Co}{ii}* troughs, \ion{Co}{ii}* $\lambda2260$ and $2286$, from which we have determined a lower limit of \ion{Co}{ii} column density. We note the reported column densities in Table~\ref{table:columndensity} of ions with multiple energy states, such as \ion{Fe}{ii} and \ion{Ni}{ii}, which are the sum of several energy states. The column densities of both \ion{Fe}{ii} and \ion{Ni}{ii} are dominated by their resonance states. We also note that the \ion{Fe}{iii} transitions are marked in Figure~\ref{fig:spectrum} for completion, but they do not have associated discernible troughs. For the photoionization analysis to be described in Section~\ref{subsec:nvu}, we added $20\%$ of the column densities to their errors in quadrature to take into account systemic uncertainties, such as that of the modeling of the continuum, following the methodology of \citet{2018ApJ...858...39X}. We note that the majority of the adopted values are lower limits, largely due to saturation of the absorption lines.
\begingroup
\setlength{\tabcolsep}{10pt}
\renewcommand{\arraystretch}{1.5}
\begin{table}
\caption{Ionic column densities of the outflow of J1321-0041.}
\centering
\begin{tabular}{lcc}
\hline\hline
Ion&AOD&Adopted\\
\hline
\ion{C}{ii}           &$31.8^{+0.9}_{-0.7}$     & $>31.8_{-4.5}$\\
\ion{C}{iv}           &$16.2^{+.3}_{-.2}$     & $>16.2_{-3.2}$\\
\ion{N}{v}           &$29.6^{+.5}_{-.5}$     & $>29.6_{-5.9}$\\
\ion{Al}{ii}           &$.90^{+.02}_{-.02}$     & $>.9_{-.2}$\\
\ion{Al}{iii}           &$4.4^{+.1}_{-.1}$     & $>4.4_{-.9}$\\
\ion{Si}{ii}          &$486^{+5}_{-5}$     & $>486_{-69}$\\
\ion{Si}{iv}           &$7.0^{+.2}_{-.1}$     & $>7.0_{-1.4}$\\
\ion{S}{ii}           &$131^{+2}_{-2}$     & $>131_{-26}$\\
\ion{S}{iv}          &$44^{+21}_{-4}$     & $44^{+22}_{-8}$\\
\ion{Fe}{ii}           &$590^{+30}_{-30}$     & $590^{+110}_{-110}$\\
\ion{Ni}{ii}           &$170^{+5}_{-4}$     & $170^{+34}_{-34}$\\
\ion{Co}{ii}           &$1.16^{+.07}_{-.05}$     & $>1.2_{-0.2}$\\
\hline
\end{tabular}
\label{table:columndensity}
\tablefoot{Column densities are shown in units of $10^{14}$ cm$^{-2}$.}
\end{table}
\endgroup
\subsection{Photoionization analysis}
\label{subsec:nvu}
\begin{figure}
   \centering
   \resizebox{\hsize}{!}
        {\includegraphics{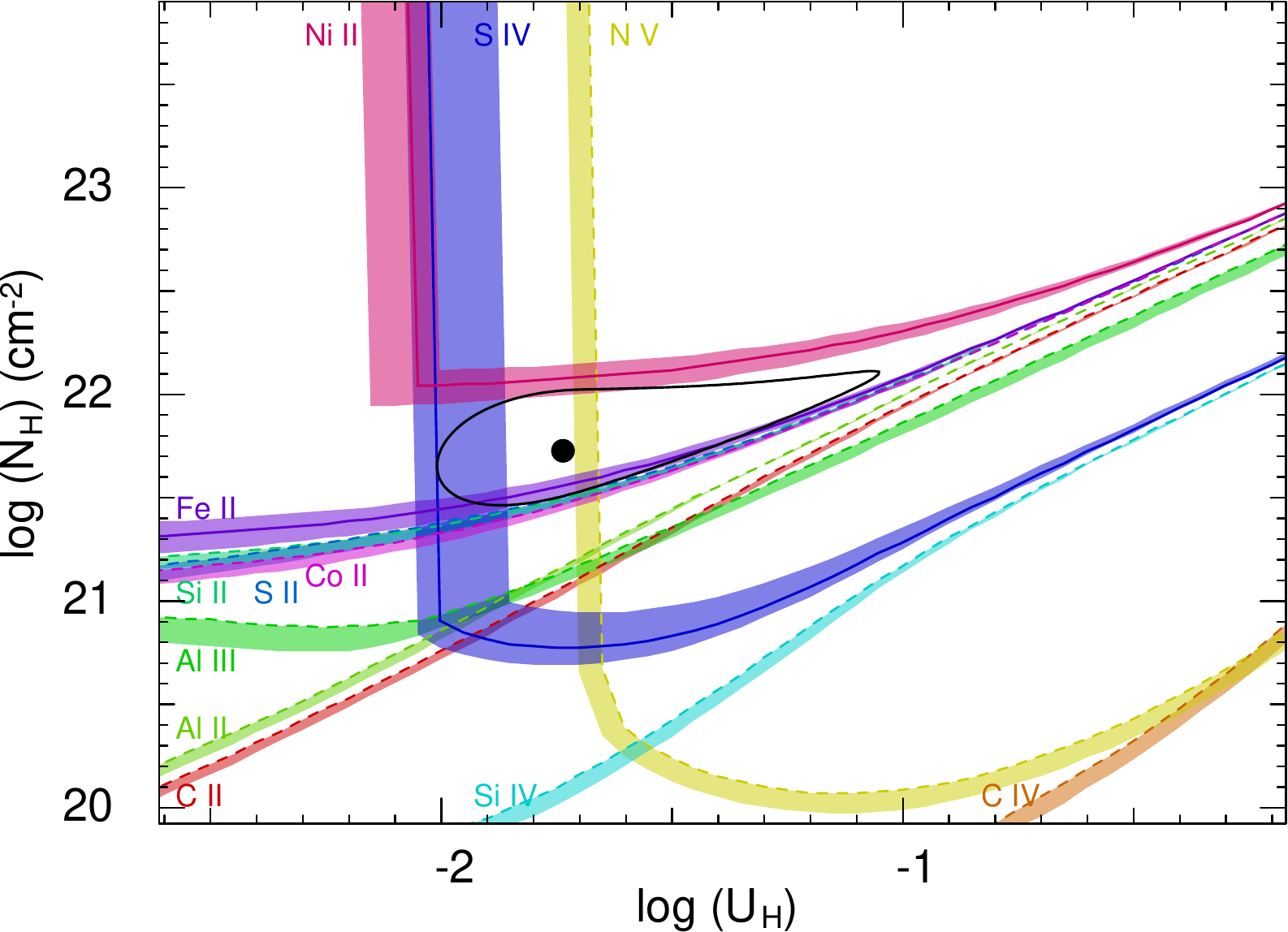}}
      \caption{Plot of hydrogen column density ($\log{N_H}$) vs. ionization parameter ($\log{U_H}$), with constraints based on the measured ionic column densities shown in Table~\ref{table:columndensity}. Measurements are shown as solid curves, while the dashed curves show lower limits. The colored shades indicate the uncertainties in the constraints of the parameters based on the uncertainties in column density. The black dot shows  the solution of $\log{N_H}$ and $\log{U_H}$ that best matches the column densities, while the black ellipse indicates the 1-$\sigma$ error.
              }
         \label{fig:nvu}
   \end{figure}

We used the spectral synthesis code \textsc{Cloudy} \citep{2017RMxAA..53..385F} to find the best fitting values of the hydrogen column density ($N_H$) and the ionization parameter ($U_H$), by comparing modeled values of ionic column densities to the measured values shown in Table~\ref{table:columndensity}. Analyses using this method have been carried out in previous works as well \citep[e.g.,][]{2022Byun,2022MNRAS.516.3778W}. As shown in Figure~\ref{fig:nvu}, \textsc{Cloudy} was used to create a grid of outflow models using a range of $N_H$ and $U_H$ values, with the measured ionic column densities serving as constraints to the parameters. Assuming solar metallicity and the spectral energy distribution (SED) of the quasar HE0238-1904 \citep[hereafter, HE0238,][]{2013MNRAS.436.3286A}, this results in the best fitting solution of $\log{N_H}=21.73^{+0.39}_{-0.26} [\text{cm}^{-2}]$ and $\log{U_H}=-1.74^{+0.69}_{-0.27}$. The SED of HE0238-1904 is the best empirically determined SED in the extreme UV, which most of the ionizing photons come from \citep{2013MNRAS.436.3286A}. We note that this solution depends on the assumption that the column density of \ion{S}{iv} is a measurement, an assumption we make as we find \ion{S}{iv} to be relatively unsaturated. Setting the column density of \ion{S}{iv} to be a lower limit results in an unbound lower limit of $U_H$. We note that the \ion{S}{iv} trough is from the SDSS spectrum, as it was outside the wavelength range of the SQUAD spectrum.
\subsection{Electron number density}
\label{subsec:ne}
\begin{figure}
   \centering
   \resizebox{\hsize}{!}
        {\includegraphics{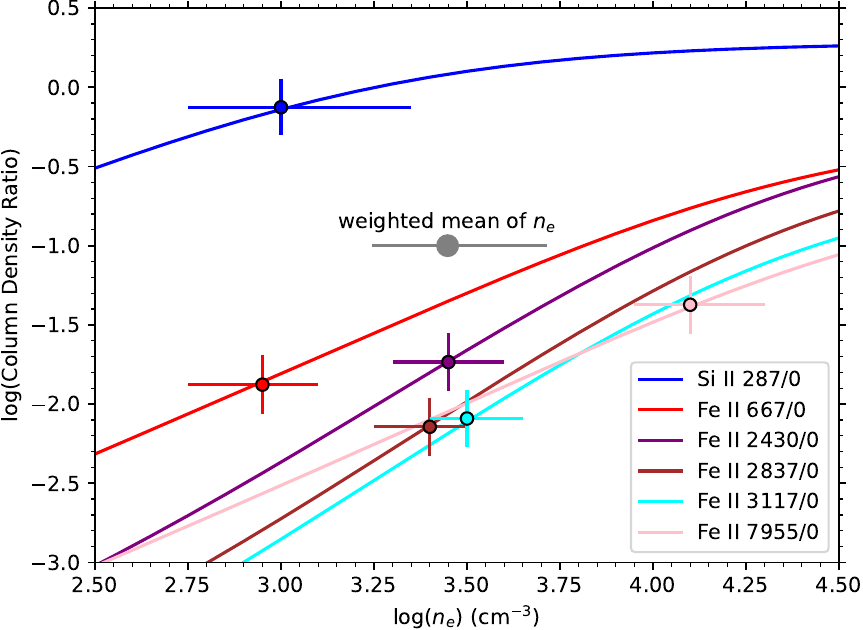}}
      \caption{Measured column density ratios of different energy states of \ion{Si}{ii} and \ion{Fe}{ii} versus electron number density ($\log{n_e}$), calculated using \textsc{Chianti}\citep{1997A&AS..125..149D,Dere_2019}. The color-coded curves are the theoretical relationships between ratio vs. $n_e$, while the dots indicate the measured ratios and associated $\log{n_e}$ values. The weighted mean of $\log{n_e}=3.45^{+0.26}_{-0.20}$ is indicated with a gray dot.
              }
         \label{fig:ratios}
   \end{figure}

Assuming collisional excitation (an assumption verified by \textsc{Cloudy} simulations), an outflow's electron number density can be found from the column density ratios between different energy states of ions \citep[e.g.,][]{2009ApJ...706..525M}. We used the troughs of \ion{Si}{ii} and \ion{Fe}{ii} that we found to be relatively unsaturated and free of contamination (see bottom panel of Figure~\ref{fig:vcut}) and overlaid the ratios between resonance and excited states of the ions with the relation between column density ratio and $n_e$ calculated using the \textsc{Chianti} atomic database \citep[version 9.0.1,][]{1997A&AS..125..149D,Dere_2019}. The \ion{Si}{ii} states were those with energies of $E(\text{cm}^{-1})=0, 287$ and those of \ion{Fe}{ii} had $E(\text{cm}^{-1})=0, 667, 2430, 2837, 3117, 7955$. The resulting $\log{n_e}$ values ranged from 2.9 to 4.2 (see Figure~\ref{fig:ratios}). We calculated the weighted mean of $\log{n_e}$, following the linear method described by \citet{2003sppp.conf..250B}. For the uncertainty, we used:
\begin{equation}
    \Delta \log{n_e}=\frac{|\log{n_e}_{,ex}-\langle\log{n_e}\rangle|}{\sqrt{N}}
,\end{equation}
where $\langle\log{n_e}\rangle$ is the weighted mean, $\log{n_e}_{,ex}$ is the maximum (minimum) measured $\log{n_e}$ used when calculating the upper (lower) error, and $N$ is the number of measured $\log{n_e}$ values. This resulted in $\log{n_e}=3.45^{+0.26}_{-0.20}$. We note that the range of $\log{n_e}$ values is within 2-$\sigma$ of the weighted mean.\par
\begin{figure}
    \centering
    \resizebox{\hsize}{!}
    {\includegraphics{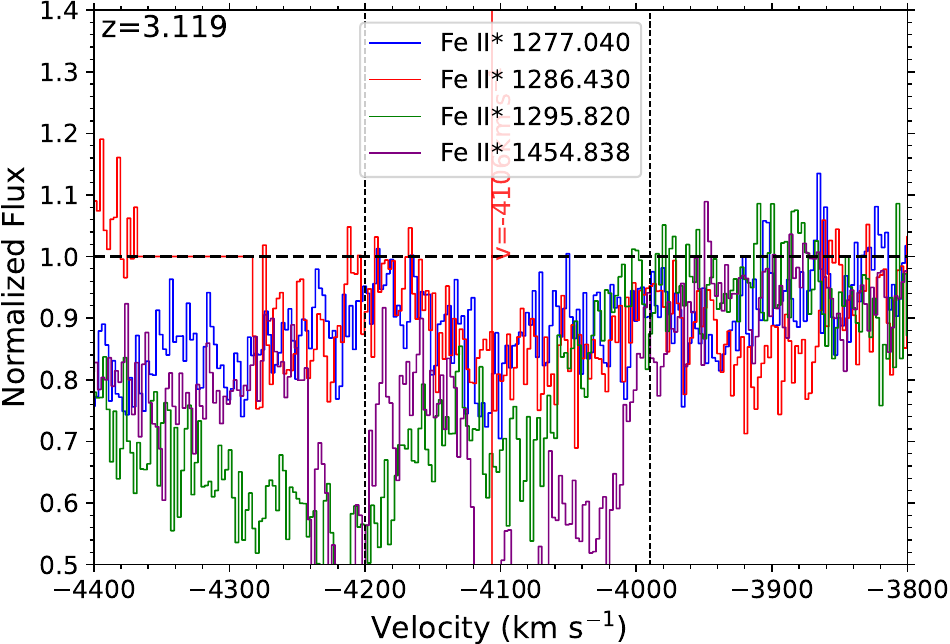}}
    \caption{\ion{Fe}{ii}* absorption troughs that have been excluded from the calculation of $\log{n_e}$.}
    \label{fig:feII_unused}
\end{figure}

We were also able to find troughs of \ion{Fe}{ii}* of energies $E=385, 13474, 22637$, and 27620, which had smaller oscillator strengths than the troughs that had been used for this analysis (See Table~\ref{table:oscillatorstrengths}). Despite the smaller oscillator strength values, the troughs are as deep as (or even deeper than) the troughs that have been used for the number density calculation (see bottom panels of Figures~\ref{fig:vcut} and \ref{fig:feII_unused} for comparison.) This suggests that the troughs may have been contaminated by unidentified absorption features, making them unreliable indicators of \ion{Fe}{ii}* column density. Thus, we have excluded them from this calculation.
\begingroup
\setlength{\tabcolsep}{10pt}
\renewcommand{\arraystretch}{1.5}
\begin{table}
\caption{Oscillator strengths of \ion{Fe}{ii} lines found in the spectrum.}
\centering
\begin{tabular}{lccc}
\hline\hline
Energy (cm$^{-1}$)&$\lambda$ (\AA)&f\\
\hline

0 & 1901.773&$6.0\times10^{-5}$\\
667 &1628.160 &$1.8\times10^{-2}$\\
2430 & 1623.093&$8.8\times10^{-3}$\\
2837 & 1570.245&$4.0\times10^{-2}$\\
3117 & 1726.393&$2.2\times10^{-2}$\\
7955 & 1562.270&$5.2\times10^{-2}$\\
\hline
385 & 1277.040&$2.7\times10^{-6}$\\
13474 & 1286.430&$3.1\times10^{-4}$\\
22637 & 1295.820&$4.0\times10^{-5}$\\
27620 & 1454.838&$2.1\times10^{-3}$\\

\hline
\end{tabular}
\tablefoot{The transitions  used in the analysis and those that are not are divided by the horizontal line.}
\label{table:oscillatorstrengths}
\end{table}
\endgroup
\section{Results}
\label{sec:results}
With $N_H$, $U_H$, and $n_e$ found (see Table~\ref{table:energetics}), we could determine the distance of the outflow from the source based on the definition of $U_H$:
\begin{equation}
    U_H \equiv \frac{Q_H}{4\mathrm{\pi}R^2 n_H c}
,\end{equation}
where $Q_H$ is the emission rate of Hydrogen ionizing photons, $R$ is the outflow's distance from the quasar, and $n_H$ is the Hydrogen number density. Following the approximation of $n_e\approx1.2 n_H$ for highly ionized plasma \citep{2006agna.book.....O}, we could calculate the value of $R$ once $Q_H$ was found.\par
In order to find $Q_H$, and by extension, the bolometric luminosity $L_{Bol}$, we scaled the SED of HE0238 to match the continuum flux at observed wavelength $\lambda=5850\AA$, $F_\lambda = 8.25\pm0.64\times10^{-17}\text{ erg s}^{-1}\text{ cm}^{-2}$. We then integrated the SED for energies above 1 Ryd, resulting in $Q_H=9.6\pm0.7\times10^{56}\text{ s}^{-1}$, and $L_{Bol}=1.72\pm0.13\times10^{47}\text{ erg s}^{-1}$. The resulting outflow distance was $R=2.5^{+1.0}_{-1.4}$ kpc.\par

\begingroup
\setlength{\tabcolsep}{10pt}
\renewcommand{\arraystretch}{1.5}
\begin{table}
\caption{Physical parameters of the outflow of J1321-0041.}
\centering
\begin{tabular}{lc}
\hline\hline
Parameter&Value\\
\hline
$\log{U_H}$         &$-1.74^{+0.69}_{-0.27}$\\
$\log{N_H}$ [cm$^{-2}$]         &$21.73^{+0.39}_{-0.26}$\\
$\log{n_e}$ [cm$^{-3}$]          &$3.45^{+0.26}_{-0.20}$\\
$R$ (pc)          &$2500^{+1000}_{-1400}$\\
$\dot{M}$ ($M_\odot$ yr$^{-1}$)          &$1600^{+2600}_{-1000}$\\
$\dot{E}_k$ ($10^{45}$ erg s$^{-1}$)          &$8.4^{+13.7}_{-5.4}$\\
$\log{\dot{E}_k}$ [erg s$^{-1}$]          &$45.92^{+0.42}_{-0.45}$\\
$\dot{E}_k/L_{Bol}$ (\%)          &$4.8^{+8.0}_{-3.1}$\\
\hline
\end{tabular}
\label{table:energetics}
\tablefoot{$n_e$ calculation has been done with \textsc{Chianti}, assuming temperature T=10,000 K.}
\end{table}
\endgroup
\section{Discusssion}
\label{sec:discussion}
\subsection{AGN Feedback Contribution}
\label{subsec:agnfeedback}
An outflow's kinetic luminosity must be at least $\sim0.5\%$ of the quasar's Eddington luminosity in order to contribute to AGN feedback \citep{2010MNRAS.401....7H,2020MNRAS.499.1522M}. Finding this ratio requires both the kinetic luminosity, $\dot{E}_k$, and Eddington luminosity, $L_{Edd}$. A quasar's $L_{Edd}$ value is typically found by measuring the width of an emission feature, such as that of \ion{C}{iv} or \ion{Mg}{ii}, and estimating the mass of the black hole \citep{2006ApJ...641..689V,2017MNRAS.465.2120C,2019ApJ...875...50B}. However, as can be seen in Figure~\ref{fig:spectrum}, the spectrum of J1321-0041 lacks a prominent emission feature which we could use to make such an estimate. Therefore, we used $L_{Bol}$ as described in Section~\ref{sec:results} as a substitute metric, assuming that $L_{Edd}\approx L_{Bol}$, as this has been the case for several quasars that have been analyzed via this method \citep[e.g.,][]{2022Byun,10.1093/mnras/stac2638}. We advise caution in taking this assumption at face value, as it has shown to not always be the case for FeLoBALQs \citep[e.g.][]{2022ApJ...935...92L}.We assumed that the geometrical shape of the outflow was that of an incomplete spherical shell, and found the mass flow rate:
\begin{equation}
    \dot{M}\simeq 4\mathrm{\pi}\Omega R N_H \mu m_p v
,\end{equation}
where $\Omega=0.2$ is the global covering factor, $\mu=1.4$ is the mean atomic mass per proton, $m_p$ is proton mass, and $v$ is outflow velocity \citep{Borguet2012,2013ApJ...762...49B}. The global covering factor 0.2 is adopted based on the $\sim20\%$ ratio of quasars in which \ion{C}{iv} BALs are found \citep{2003AJ....125.1784H}. Despite the relative rarity of low-ionized BALs with troughs such as \ion{Si}{ii} or \ion{C}{ii}, it is likely that quasars with such LoBALs are BALQSOs at a specific line of sight (See \citet{2010ApJ...709..611D} for a full discussion). We then calculated the kinetic luminosity ($\dot{E}_k=\frac{1}{2}\dot{M}v^2$), resulting in $\dot{E}_k=8.4^{+13.2}_{-5.3}\times10^{45}\text{ erg s}^{-1}$. The ratio between the kinetic luminosity and bolometric luminosity is $\dot{E}_k/L_{Bol}=4.8^{+7.7}_{-3.1}\%$, which would be sufficient to contribute to AGN feedback (see Table~\ref{table:energetics} for a full summary of the parameters).
\subsection{Comparison with other outflows}
\label{subsec:other}
The $n_H$ value of the J1321-0041 outflow was found mainly by examining the ratios between the \ion{Fe}{ii} excited and resonance state column densities. This has been done in past studies as well \citep[e.g.,][]{2008ApJ...672..108D,10.1093/mnras/stac2194}. The $n_e$ values from the different ratios agree within $\sim0.5$ dex with the mean (see Figure~\ref{fig:ratios}) and combined with the previous outflow analyses, this demonstrates that \ion{Fe}{ii} is an effective probe for $n_H$ and outflow distance.\par
In previous analyses of FeLoBALs based on high resolution VLT/UVES data, the distance $R$ has been measured to be within the range of $1\text{ kpc} \lesssim R \lesssim 67\text{ kpc}$ \citep[e.g.,][]{2022Byun,10.1093/mnras/stac2194,2022MNRAS.516.3778W}. The distance of the J1321-0041 outflow from its source is $R\approx2500$ pc, which is well within this range.\par
As previously mentioned, there have been studies analyzing extreme FeLoBALs in the past. \citet{2021MNRAS.506.2725X} analyzed the physical conditions of the FeLoBAL of the quasar Q0059-2735, which had a \ion{C}{iv} width of $\sim25,000\text{ km s}^{-1}$, roughly twice as wide as the \ion{C}{iv} width of J1321-0041 ($\sim10,000\text{ km s}^{-1}$) found in its SDSS spectrum. \citet{2022ApJ...937...74C} studied a larger sample of FeLoBALs, some of which have extreme blending in their absorption troughs. The presence of unblended absorption features in the J1321-0041 spectrum enabled us to conduct a precise analysis of the physical parameters of the outflow system.\par

\section{Summary and conclusion}
\label{sec:conclusion}
We have identified an FeLoBAL outflow system in the VLT/UVES spectrum of the quasar SDSS J1321-0041. Through the measurement of ionic column densities and photoionization analysis, we determined the Hydrogen column density and ionization parameter of the outflow (see Figure~\ref{fig:nvu} and Table~\ref{table:energetics}).\par
The presence of \ion{Fe}{ii} and \ion{Si}{ii} absorption trough enabled us to find the electron number density $n_e$ by using \textsc{Chianti} to relate their column density ratios to estimates of $n_e$. This allowed us to determine the outflows distance from its central source, kinetic luminosity, and the ratio between kinetic luminosity and the quasar's bolometric luminosity ($\dot{E}_k/L_{Bol}=4.8^{+8.0}_{-3.1}$). Assuming $L_{Bol}\approx L_{Edd}$, the value of $\dot{E}_k$ is above the required threshold to contribute to AGN feedback.

\begin{acknowledgements}
We acknowledge support from NSF grant AST 2106249, as well as NASA STScI grants AR-15786, AR-16600, and AR-16601.
\end{acknowledgements}

\bibliographystyle{aa}
\bibliography{J1321}

\end{document}